\documentclass[doublecol]{epl2}

\title{Controllable photo-induced spin and valley filtering in silicene}

\author{Y. Mohammadi\inst{1} \and B. Arghavani Nia\inst{2}}

\institute{
  \inst{1} Young Researchers and Elite Club, Kermanshah
Branch, Islamic Azad University, Kermanshah, Iran \\
  \inst{2} Department of Physics, Kermanshah Branch
Islamic Azad University, Kermanshah, Iran}


\abstract{We study ballistic transport of Dirac electrons through
a strip in silicene, when the strip is exposed to off-resonant
circularly polarized light and an electric filed applied
perpendicular to the silicene plane. We show that the conductance
through the strip is spin- or/and valley-polarized. This can be
explained by spin-valley coupling in silicene, and modification of
its band structure through virtual absorption/emission processes
and also by the perpendicular electric field. The spin- (valley-)
polarization can be enhanced by tuning the light intensity and the
value of the perpendicular electric field, leading to perfect spin
(valley) filtering for certain of their values. Further, the spin
(valley) polarization can be inverted by reversing the
perpendicular electric field (by reversing the perpendicular
electric field or reversing the circular polarization of the light
irradiation). The conditions necessary for the fully valley
polarization is determined.}
\begin{document}
\maketitle

\section{Introduction}

Silicene, a monolayer of silicon atoms arranged in a honeycomb
lattice structure, has been synthesized
recently\cite{Vogt1,Lalmi1,Chen1,Meng1}. Due to large ionic radius
of silicon atoms, the honeycomb lattice structure is puckered such
that the A and B sublattices are shifted vertically with respect
to each other and sit in two parallel planes with a separation of
$0.46~nm$\cite{Drummond1,Ni1}. Due to the puckered structure,
which results in a large spin-orbit interaction\cite{Liu1}, the
low energy dynamic near the Dirac points in the hexagonal
Brillouin zone of silicene is dominated by a massive Dirac
Hamiltonian\cite{Liu1}, with a mass which could also be tuned via
an electric filed applied perpendicular to the silicene
plane\cite{Drummond1,Ezawa1}. These novel features donate many
attractive properties to silicene
\cite{Drummond1,Ezawa1,Ezawa2,Ezawa3,An1,Tahir1,Tsai1,Tabert1,Pan1}

In silicene, valley and spin degrees of freedom have been coupled
via a spin-orbit interaction\cite{Liu1} which is large compared
with that in graphene\cite{Min1}. This can lead to detectable
spin- or/and valley-polarized transport in silicene, if the spin
or/and valley degeneracies of its band structure are
lifted\cite{Tabert1,Mohammadi1}. Further, silicon has a long
spin-coherence length\cite{Sanvito1} and spin-diffusion
time\cite{Huang1,Wang1}. Motivated by these facts, recently
several groups have investigated ballistic transport of Dirac
fermions in ferromagnetic silicene
junctions\cite{Yokoyama1,Vargiamidis1,Missault1}. They studied the
influence of electric and exchange fields on the ballistic
transport across single\cite{Yokoyama1,Vargiamidis1},
double\cite{Vargiamidis1} and arrays\cite{Missault1} of
ferromagnetic (FM) barriers. They reported novel results such as
perfect spin and/or valley polarization and tunable transport gap
which are electrically controllable.

In this letter, we propose a scheme to employ off-resonant
circularly polarized light to achieve electrically/optically
controllable nearly perfect spin- and valley-filtering in
silicene. Our motivation to propose this scheme is the development
of new experimental probes\cite{Wang2,Sie1} which make it possible
to access non-equilibrium effects arising from off-resonant light
irradiation. In our proposed scheme, a strip in a silicene plane
is exposed to off-resonant circularly polarized light. Hence, the
band structure inside the strip is modified through virtual photon
absorption/emission processes, leading to a new energy band which,
due to the spin-valley coupling in silicene, is spin-polarized. If
a perpendicular electric field is applied, the band structure
becomes valley-polarized too. Such a light-irradiated strip can
act as a spin- and/or valley-dependent barrier, leading to nearly
perfect spin or/and valley filtering for certain values of the
light intensity and the perpendicular electric field.

\section{Model Hamiltonian}
\begin{figure}
\onefigure[width=7cm,height=5.0cm,angle=0]{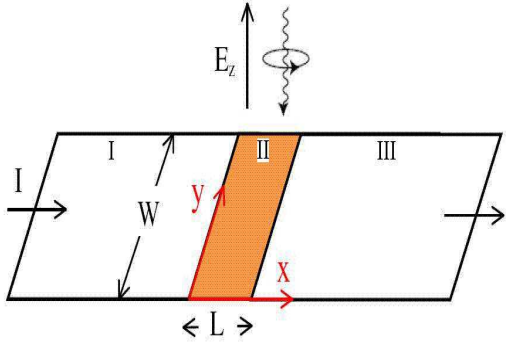}
\caption{Schematic of a silicene plane in which a strip (the
orange region), is exposed to a perpendicular electric field and
off-resonant circularly polarized light.} \label{Fig01}
\end{figure}

We study ballistic transport of Dirac fermions across a strip in a
silicene plane, when the strip is exposed to off-resonant
circularly polarized light and a perpendicular electric filed. Let
us take x-axis perpendicular to the edges of the strip (the
interfaces) and y-axis along them (see fig. \ref{Fig01}). The
interfaces are located at $x=0$ and $x=L$. We also assume the
translational invariance along y-axis satisfied in the limit of
large $W$ (the width of the silicene plane). We restrict our
consideration to $W/L\gg1$ limit, in which the effects of the
microscopic details of the upper and lower edges of the silicene
plane on the electron transport become
insignificant\cite{Tworzydlo1}.

The low energy excitations in silicene, which occur in vicinity of
the Dirac points ($\mathbf{K}$ and $\mathbf{K}^{'}$), are
dominated\cite{Liu1} by a $2\times2$ Hamiltonian matrix as
\begin{eqnarray}
H^{\eta,s_{z}}=\hbar v_{F}(k_{x}\tau_{x}-\eta k_{y}\tau_{y})-\eta
s_{z}\Delta_{so}\tau_{z}, \label{eq01}
\end{eqnarray}
acting in the sublattice pseudospin space. The first part of the
Hamiltonian is the Dirac Hamiltonian arising from the nearest
neighbor transfer energy with $\eta=+$ ($\eta=-$) for $\mathbf{K}$
($\mathbf{K}^{'}$). In this term $v_{F}=\frac{\sqrt{3}ta}{2\hbar}$
is the Fermi velocity with $t=1.6~eV$ and $a=0.386~nm$ being the
nearest-neighbor transfer energy and the lattice constant of
silicene respectively. $\tau_{i}$ ($i=x,y,z$) are the Pauli
matrixes and $\mathbf{k}=(k_{x},k_{y})$ is the two dimensional
momentum measured from the Dirac points. The second term is the
Kane-Mele term\cite{Kane1} for the intrinsic spin-orbit coupling,
in which $\Delta_{so}=3.9 meV$\cite{Liu1} is the spin-orbit
coupling and $s_{z}$ index referred to two spin degrees of freedom
with $s_{z}=+1$ and $s_{z}=-1$ for up and down spin degrees of
freedom respectively.

As mentioned above, the strip is exposed to a perpendicular
electric filed and off-resonant circularly polarized light. Due to
the buckled structure of silicene, applying the perpendicular
electric filed, $E_{z}$, causes a staggered sublattice potential
as\cite{Ezawa1} $\Delta_{z}\tau_{z}$ where $\Delta_{z}=e \ell
E_{z}$. The circularly polarized light is described by an
electromagnetic potential as
\begin{eqnarray}
\mathbf{A}(t)=(\pm A \sin\Omega t,A \cos\Omega t),\label{eq02}
\end{eqnarray}
where $\Omega$
is the frequency of the light and plus (minus) sign corresponds to
the right (left) circulation. The gauge potential is periodic in
time, $A(t+T)=A(t)$, with the time periodicity $T=2\pi/\Omega$.
The light intensity can be characterized by a dimensionless
parameter as $\mathcal{A}=eaA/\hbar$, where $e$ is the electron
charge. To capture the effects of the light irradiation, one can
apply the minimal substitution, $\hbar k_{i}\longrightarrow
P_{i}\equiv \hbar k_{i}+e \mathbf{A}_{i}$, and use Floquet theory
\cite{Kitgawa1,Oka1,Inoue1}. In this paper, we focuss on the
off-resonant frequency regime which is satisfied when $\hbar
\Omega\gg t$ in our calculation. In this regime the light does not
directly excite the electrons and instead effectively modifies the
electron band structure through virtual photon absorption/emission
processes. In the off-resonant frequency regime and in the limit
of small light intensity, $\mathcal{A}\ll 1$, the influence of
such off-resonant light on the electron band structure is well
described by a static effective Hamiltonian\cite{Ezawa3,Kitgawa1}
as
\begin{eqnarray}
\Delta
H^{\eta,s_{z}}=-\frac{[H^{\eta,s_{z}}_{+1},H^{\eta,s_{z}}_{-1}]}{\hbar
\Omega}+\mathcal{O}(\mathcal{A}^{4}),\label{eq03}
\end{eqnarray}
where $H^{\eta,s_{z}}_{m}=(1/T)\int_{0}^{T}e^{-im\Omega
t}H^{\eta,s_{z}}(t)dt$. It is easy to show that
$[H^{\eta,s_{z}}_{+1},H^{\eta,s_{z}}_{-1}]= \pm \eta(\hbar
v_{F})^{2} \mathcal{A}^{2} \tau_{z}/a^{2}$ with $+(-)$ for right
(left) circulation. Hence the low energy dynamics in silicene, in
the presence of off-resonant circularly polarized light
irradiation and an electric filed applied perpendicular to the
silicene plane, are dominated by an effective Hamiltonian as
\begin{eqnarray}
H^{\eta,s_{z}}_{eff}=H^{\eta,s_{z}}+\Delta_{z}\tau_{z}-\eta\Delta_{\Omega}\tau_{z},\label{eq04}
\end{eqnarray}
acting in the sublattice pseudospin space where
$\Delta_{\Omega}=\pm (\hbar v_{F})^{2} \mathcal{A}^{2}/a^{2}\hbar
\Omega $ with $+(-)$ for right (left) circulation. The
corresponding energy bands are obtained by diagonalizing the
Hamiltonian, Eq. \ref{eq04}, which are given by
\begin{eqnarray}
\varepsilon_{\eta,s_{z}}=\nu \sqrt{(\hbar
v_{F}|\mathbf{k}|)^{2}+\Delta^{2}},\label{eq05}
\end{eqnarray}
where $\Delta=\Delta_{z}-\eta
s_{z}\Delta_{so}-\eta\Delta_{\Omega}$ and  $\nu=+(-)$ denotes to
the conduction (valance) band. The low energy bands, in the
regions I and III, are obtained from Eq. \ref{eq05} by just
setting $\Delta_{z}=0$ and $\Delta_{\Omega}=0$.

The wave functions in the different regions (see fig. \ref{Fig01})
can be written in terms of the incident and reflected waves. In
the regions I and III , they are
\begin{eqnarray} \psi^{\eta,s_{z}}_{\nu}&=&
e^{ik_{y}y}[\frac{e^{ik_{x}x}}{\sqrt{2\chi}}\left(
\begin{array}{c}
\sqrt{\chi-\nu \Delta_{so}}  \\
\nu e^{-i\eta \theta}\sqrt{\chi+\nu \Delta_{so}}
\end{array}
\right)  \nonumber \\
&+&  r_{\eta,s_{z}}\frac{e^{-ik_{x}x}}{\sqrt{2\chi}}\left(
\begin{array}{c}
\sqrt{\chi-\nu \Delta_{so}}  \\
\nu e^{-i\eta(\pi-\theta)}\sqrt{\chi+\nu \Delta_{so}}
\end{array}
\right)] ,\label{eq06}
\end{eqnarray}
in the region I and
\begin{eqnarray} \psi^{\eta,s_{z}}_{\nu}=
t_{\eta,s_{z}}\frac{e^{i(k_{x}x+k_{y}y})}{\sqrt{2\chi}}\left(
\begin{array}{c}
\sqrt{\chi-\nu \Delta_{so}}  \\
\nu e^{-i\eta \theta}\sqrt{\chi+\nu \Delta_{so}}
\end{array}
\right),\label{eq07}
\end{eqnarray}
in the region III, where $\nu=+(-)$ denotes to the conduction
(valance) band, $\chi=\sqrt{(\hbar
v_{F}|\mathbf{k}|)^{2}+\Delta_{so}^{2}}$,
$\theta=\tan^{-1}(k_{y}/k_{x})$ is the angle of incidence and
$r_{\eta,s_{z}}$, and $t_{\eta,s_{z}}$ are the reflection and the
transmission coefficients respectively.

The corresponding wave function, in the strip (region II), is
given by
\begin{eqnarray} \psi^{\eta,s_{z}}_{\nu}&=&
e^{iq_{y}y}[a_{\eta,s_{z}}e^{iq_{x}x}\left(
\begin{array}{c}
\sqrt{\chi^{'}-\nu \Delta}  \\
\nu e^{-i\eta \phi}\sqrt{\chi^{'}+\nu \Delta}
\end{array}
\right)  \nonumber \\
&+& b_{\eta,s_{z}} e^{-iq_{x}x}\left(
\begin{array}{c}
\sqrt{\chi^{'}-\nu \Delta}  \\
\nu e^{-i\eta(\pi-\phi)}\sqrt{\chi^{'}+\nu \Delta}
\end{array}
\right)],\label{eq08}
\end{eqnarray}
where $\chi^{'}=\sqrt{(\hbar v_{F}|\mathbf{q}|)^{2}+\Delta^{2}}$
and $\phi=\tan^{-1}(q_{y}/q_{x})$ is refractive angle inside the
strip with $\mathbf{q}=(q_{x},q_{y})$ being the two dimensional
momentum. The coefficients $a_{\eta,s_{z}}$ and $b_{\eta,s_{z}}$
could be determined by requiring continuity of the wave functions
and their derivatives at the interfaces.

We restrict our calculations to the elastic scattering at the
interfaces. Hence, the translational invariance along y-axis
implies conversation of the transverse momentums outside the
strip, $k_{y}$, and inside it, $q_{y}$, which yields $k_{F}\sin
\theta=q_{F}\sin \phi, \label{eq09}$ where
\begin{eqnarray}
k_{F}=(k_{x}^{2}+k_{y}^{2})^{1/2}=(\hbar v_{F})^{-1}
\sqrt{\varepsilon_{F}^{2}-\Delta_{so}^{2}}, \label{eq09}
\end{eqnarray}
and
\begin{eqnarray}
q_{F}=(q_{x}^{2}+q_{y}^{2})^{1/2}=(\hbar v_{F})^{-1}
\sqrt{\varepsilon_{F}^{2}-\Delta^{2}}, \label{eq10}
\end{eqnarray}
with $k_{F}$ and $q_{F}$ being the Fermi momentum outside and
inside the strip respectively with $\varepsilon_{F}$ being the
Fermi energy. The coefficients of the incident and reflected wave
functions could be determined by requiring continuity of the wave
functions and their derivatives at the interfaces. Matching the
wave functions and their derivatives yield same results. So, we
only need to match the wave functions at the interfaces, $x=0$ and
$x=L$, which ensures the conversation of the local current at the
interfaces too. These conditions yield
\begin{eqnarray}
t_{\eta,s_{z}}=(\cos(q_{x}L)-i
\mathcal{F}(k_{F},\theta)\sin(q_{x}L))^{-1}, \label{eq11}
\end{eqnarray}
for the transmission coefficient through the strip where
\begin{eqnarray}
q_{x}=k_{F}\sqrt{\frac{\varepsilon_{F}^{2}-\Delta^{2}}{\varepsilon_{F}^{2}-\Delta_{so}^{2}}-\sin^{2}\theta},
\label{eq12}
\end{eqnarray}
and
\begin{eqnarray}
\mathcal{F}(k_{F},\theta)=\frac{k_{x}^{2}\varepsilon_{b}^{2}+q_{x}^{2}\varepsilon_{l}^{2}
+k_{y}^{2}(\varepsilon_{b}-\varepsilon_{l})^{2}}{2\varepsilon_{l}\varepsilon_{b}k_{x}q_{x}},
\label{eq13}
\end{eqnarray}
with $k_{x}=k_{F}\cos \theta$,
$\varepsilon_{l}=\varepsilon_{F}+\Delta_{so}$ and
$\varepsilon_{b}=\varepsilon_{F}+\Delta$. Here we remark a point
worth mentioning. Note that for $q_{x}L=n\pi$, with $n$ being an
integer number, the transmission is perfect, called
Fabry-P\'{e}rot resonance. This occurs when $n$ times half the
wavelength of the wave inside the barrier is equal to the length
of the strip.

Spin- and valley-resolved conductance in the Landaur-B\"{u}ttiker
formalism are given by
\begin{eqnarray}
G_{\eta,s_{z}}=\frac{G_{0}}{2}\int_{-\pi/2}^{\pi/2}T_{\eta,s_{z}}(k_{F},\theta)\cos
\theta d\theta , \label{eq14}
\end{eqnarray}
where $G_{0}=\frac{e^{2}}{h}\frac{k_{F}W}{\pi}$ and $W$ is the
width of the silicene sheet. To explore the spin/valley
polarization through the strip, we define the spin and valley
polarization as
$P_{s}=(G_{\uparrow}-G_{\downarrow})/(G_{\uparrow}+G_{\downarrow})$
and $P_{v}=(G_{K}-G_{K^{'}}/(G_{K}+G_{K^{'}})$ respectively where
$G_{s_{z}}=\sum_{\eta}G_{\eta,s_{z}}$ and
$G_{\eta}=\sum_{s_{z}}G_{\eta,s_{z}}$. So, $P_{s}=+1(-1)$
corresponds to a current which is carried entirely by spin-up
(spin-down) electrons, while $P_{v}=+1(-1)$ corresponds to a
current that is carrier by the electrons localized completely in
the $\mathbf{K}(\mathbf{K}^{'})$ valley. The total charge
conductance is obtained by summing $G_{\eta,s_{z}}$ over $s_{z}$
and $\eta$.

\section{Results and Discussion}

In this section we present our results for conductance, spin and
valley polarizations and use them to discuss off-resonant
photo-induced spin- and valley-filtering in silicene. We first
discuss it in the absence of the perpendicular electric field
($\Delta_{z}=0$) and then in its presence ($\Delta_{z}\neq0$).

\begin{figure}
\onefigure[width=7.0cm,height=5.0cm,angle=0]{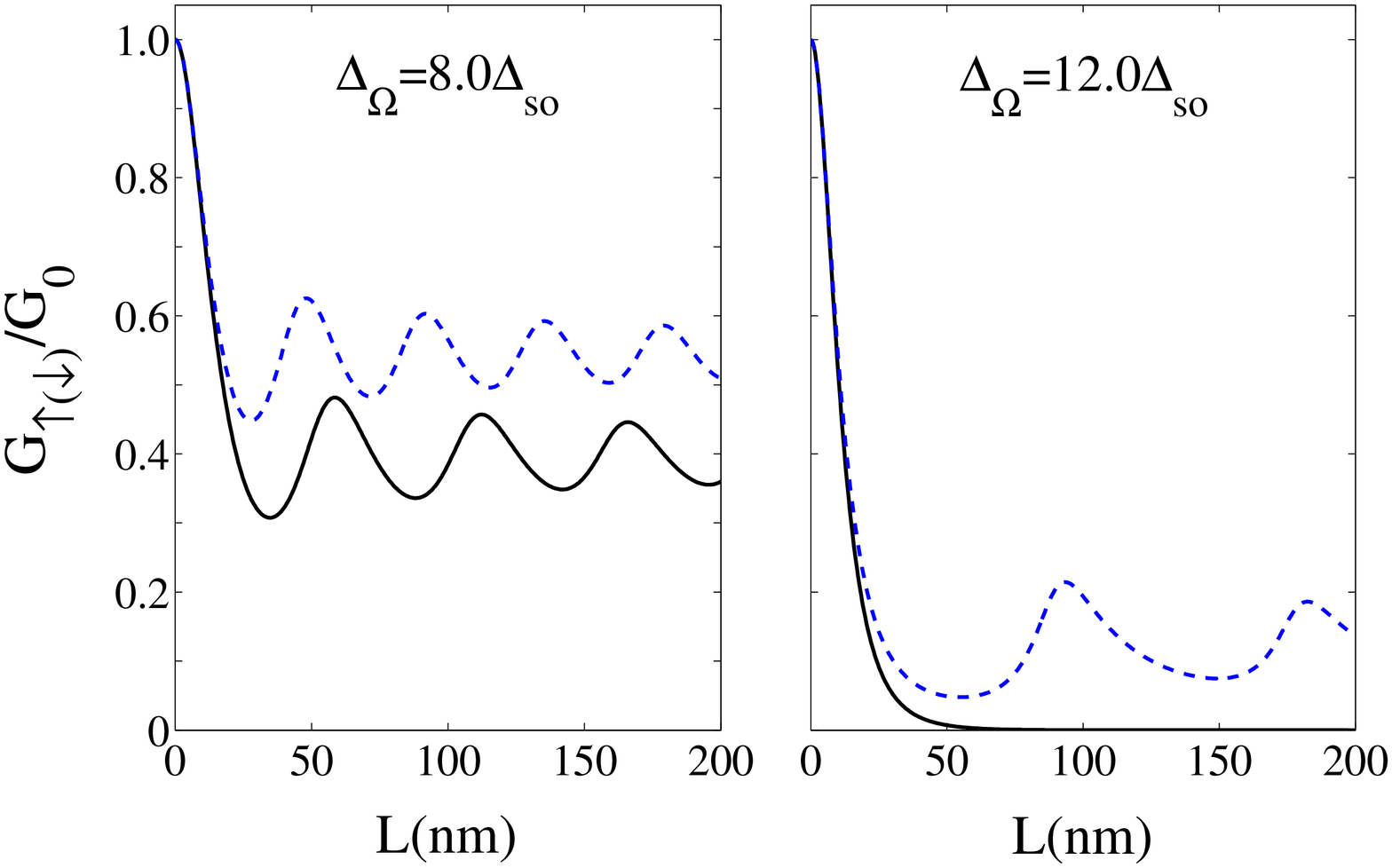}
\caption{Spin-resolved conductances, $G_{\uparrow}$ (solid black
curves) and $G_{\downarrow}$ (dashed blue curves), as a function
of $L$ for $\Delta_{\Omega}=8.0\Delta_{so}$ (right panel) and
$\Delta_{\Omega}=12.0\Delta_{so}$ (left panel). The other
parameters are $\varepsilon_{F}=12.0\Delta_{so}$,
$G_{0}=\frac{e^{2}}{h}\frac{k_{F}W}{\pi}$ and $\Delta_{z}=0$.}
\label{Fig02}
\end{figure}

\begin{figure}
\onefigure[width=7.0cm,height=5.0cm,angle=0]{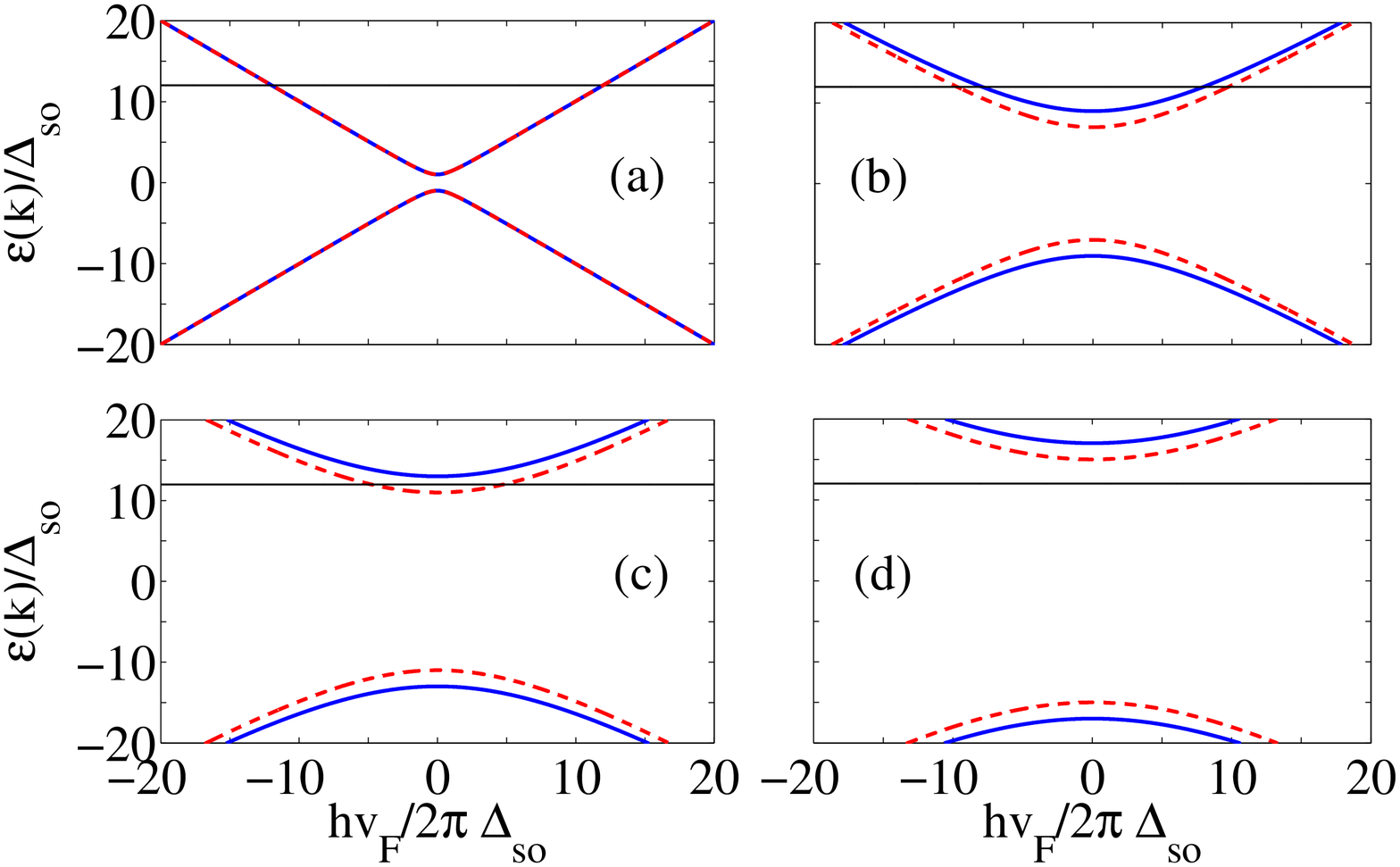}
\caption{Electronic band structure of silicene around Dirac points
as a function of $\hbar v_{F}/\Delta_{so}$ for different values of
$\Delta_{\Omega}$, (a) $\Delta_{\Omega}=0$ , (b)
$\Delta_{\Omega}=8\Delta_{so}$, (c)
$\Delta_{\Omega}=12\Delta_{so}$ and (d)
$\Delta_{\Omega}=16\Delta_{so}$. The vertical line shows the Fermi
energy $\varepsilon_{F}=12\Delta_{so}$ and $\Delta_{z}=0$.}
\label{Fig03}
\end{figure}

$\mathbf{\Delta_{z}=0}:$ In the presence of off-resonant light
irradiation, the energy bands are modified through virtual photon
absorption/emission processes. Such photon-dressed bands in
silicene, due to the spin-valley coupling, become spin-polarized.
So if the light intensity is tuned properly, the conductance
trough the strip can becomes partially or even fully
spin-polarized provided the Fermi level crosses the conduction (or
valance) bands. We have plotted in fig. \ref{Fig02} the
spin-resolved conductance through the strip, $G_{\uparrow}$ (solid
black curves) and $G_{\downarrow}$ (dashed blue curves), as a
function of $L$ for two different values of $\Delta_{\Omega}$,
$\Delta_{\Omega}=8.0\Delta_{so}$ (right panel) and
$\Delta_{\Omega}=12.0\Delta_{so}$ (left panel). The Fermi energy
for both left and right panels is
$\varepsilon_{F}=12.0\Delta_{so}$. In the left panel, both
$G_{\uparrow}$ and $G_{\downarrow}$ exhibit an oscillatory
decaying dependence on the length of the strip, without being
completely suppressed even for large $L$. This can be explained by
the band structure of the silicene inside the strip, in which the
Fermi level crosses both spin-up and spin-down electron bands (see
fig. \ref{Fig03} (b)), leading to propagating modes inside the
strip for both spin wave functions. Further, left panel of fig.
\ref{Fig02} shows a partially spin polarization which can be
enhanced by increasing $\Delta_{\Omega}$, leading to a fully spin
polarization as seen in the right panel in which $G_{\downarrow}$
decays with $L$ in a oscillatory way while $G_{\uparrow}$ is
strongly suppressed $\textit{if the length of the strip is large
enough}$. This is due to the fact that the Fermi level crosses
only the spin-down energy band around each Dirac point inside the
strip (see fig. \ref{Fig03} (c)), satisfied when $\Delta_{\Omega}$
becomes larger than a critical value,
$|\Delta_{\Omega}^{c}|=\varepsilon_{F}-\Delta_{so}$. Hence,
$q_{x}$ becomes imaginary leading to monotonically decaying the
transmission probability and consequently the spin-resolved
conductance in terms of $L$ (see Eqs. \ref{eq09}-\ref{eq12}).

The spin polarization could be inverted by reversing the
polarization of the light. Because changing circular polarization
of the light from right to left polarization or vise versa
interchanges the spin-up and spin-down electron bands in the
strip. This is clear from the fig. \ref{Fig03}, in which we have
plotted the spin polarization as a function of $\Delta_{\Omega}$
for different values of the Fermi energies,
$\varepsilon_{F}=8\Delta_{so}$, $\varepsilon_{F}=12\Delta_{so}$
and $\varepsilon_{F}=16\Delta_{so}$. This figure, with the band
structure of silicene inside the strip, can also be used to
determine the necessary condition to realize a fully spin
polarization. This condition is
$\varepsilon_{F}-\Delta_{so}<|\Delta_{\Omega}|<\varepsilon_{F}+\Delta_{so}$
and $\varepsilon_{F}>\Delta_{so}$. \textit{The results presented
in figs. \ref{Fig02} and \ref{Fig03} reveal off-resonant
photo-induced perfect spin filtering in silicene.}
\begin{figure}
\onefigure[width=7.0cm,height=5.0cm,angle=0]{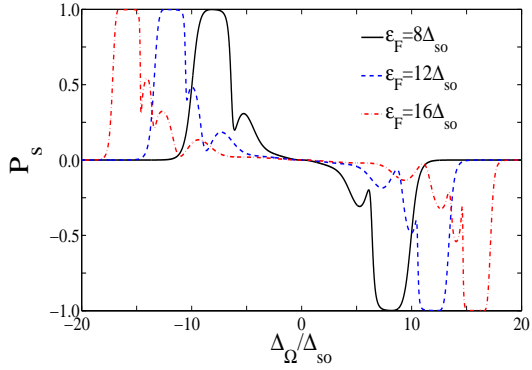}
\caption{Spin resolved conductances, $G_{\uparrow}$ (solid black
curves) and $G_{\downarrow}$ (dashed blue curves), as a function
of $\Delta_{\Omega}/\Delta_{so}$. The other parameters are $L=125
nm$, $\varepsilon_{F}=12.0\Delta_{so}$, $\Delta_{z}=0$.}
\label{Fig04}
\end{figure}
%

\begin{figure}
\onefigure[width=7.0cm,height=5.0cm,angle=0]{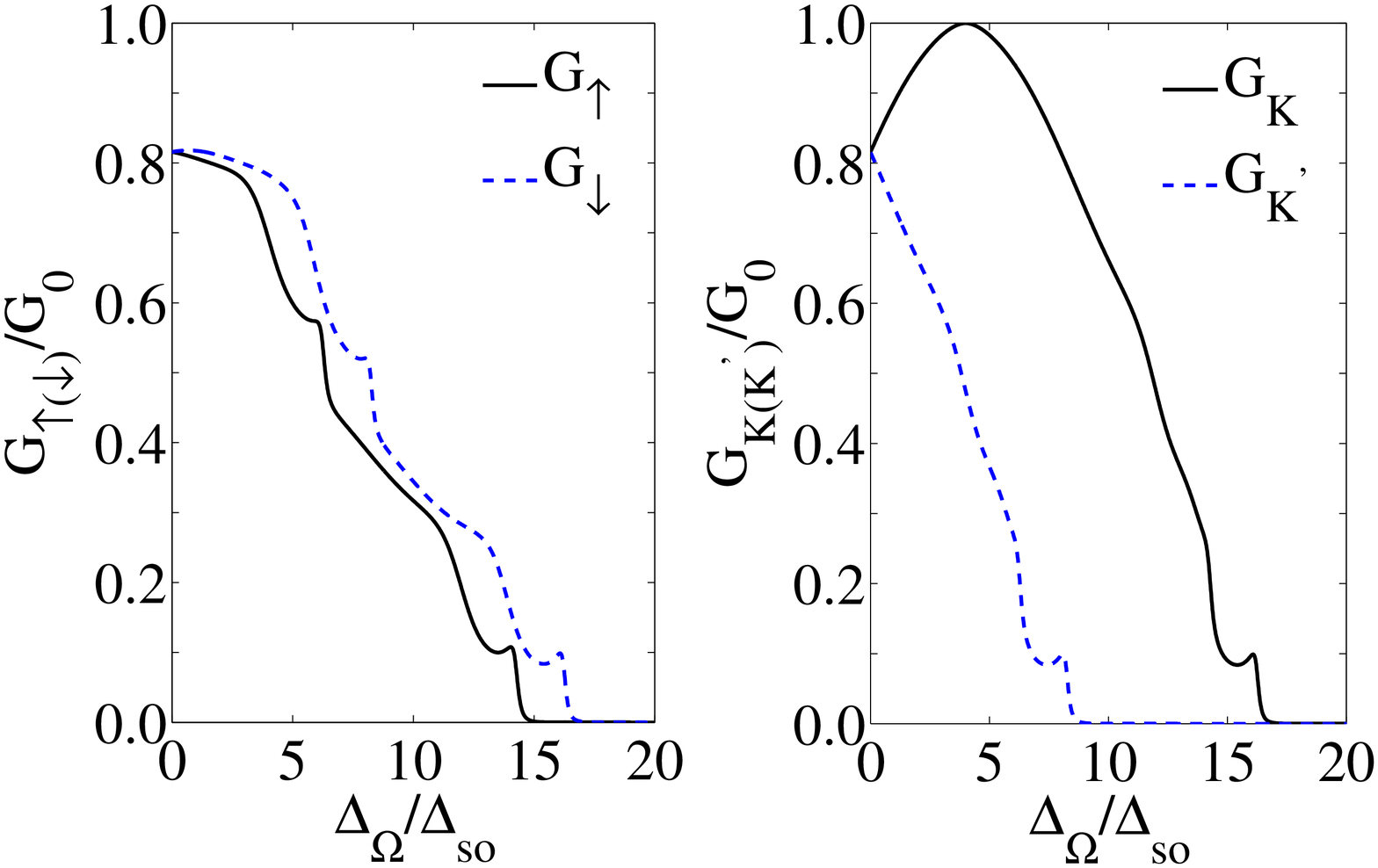}
\caption{(a) Spin-resolved conductance and (b) Valley-resolved
conductance as a function of $\Delta_{\Omega}/\Delta_{so}$. The
other parameters are $\varepsilon_{F}=12\Delta_{so}$,
$\Delta_{z}=4\Delta_{so}$ and $L=100~nm$.} \label{Fig05}
\end{figure}

$\mathbf{\Delta_{z}\neq0}:$ Applying a perpendicular electric
($\Delta_{z}\neq 0$) lifts the valley-degeneracy of the energy
bands (see Eq. \ref{eq05}). So, in addition to a spin
polarization, one can achieve a partially valley polarization. The
partially spin/valley polarization can be enhanced, giving rise to
a fully spin/valley polarization. This can be done by changing the
intensity of the light (and consequently $\Delta_{\Omega}$) for a
non-zero value of the constant $\Delta_{z}$ or vise versa. Figure
\ref{Fig05} displays the spin-resolved (left panel) and the
valley-resolved (right panel) conductances across the strip as
functions of $\Delta_{\Omega}/\Delta_{so}$ for
$\Delta_{z}=4\Delta_{\Omega}$ when
$\varepsilon_{F}=12\Delta_{\Omega}$ and $L=100~nm$. Let us first
consider the dependence of the spin polarization on
$\Delta_{\Omega}$ and $\Delta_{z}$, and appearance of a fully spin
polarization in the left panel. This can be understood by the band
structure in the strip. With changing $\Delta_{\Omega}$, the
energy bands and consequently the hight of the barrier in the
strip are changed, leading to change in the transmission
probability and consequently the charge conductance. If
$\Delta_{\Omega}$ increases such that the bottom of an energy band
approaches (leaves) the Fermi level, its carrier concentration
decreases (increases). Hence, the conductance arising from that
energy band decreases (increases). Once the Fermi level crosses
the bottom of an energy band and locates inside its energy gap,
the conductance, arising from that energy band, is suppressed
completely. So, the total conductance (the spin- or
valley-resolved conductance in fig. \ref{Fig05}) decreases
intensively. Such a decrease takes place at $\Delta_{\Omega}=7
\Delta_{so}$ ($\Delta_{\Omega}=9 \Delta_{so}$) for $G_{\uparrow}$
($G_{\downarrow}$) in left panel of fig. \ref{Fig05} where the
Fermi level crosses the bottom of $\varepsilon_{K\uparrow}$
($\varepsilon_{K\downarrow}$) band and locates in its energy gap.
This happens when
$\varepsilon_{F}=|\Delta|=|\Delta_{z}+\eta(s_{z}\Delta_{so}+\Delta_{\Omega})|$.
If $\Delta_{\Omega}$ increases further and becomes larger than a
critical value ($\Delta_{\Omega}^{c}=15\Delta_{so}$), the Fermi
level locates inside the gap of $\varepsilon_{K^{'}\uparrow}$ band
too satisfied when
$\varepsilon_{F}<|\Delta_{z}+(-)(+\Delta_{so}+\Delta_{\Omega})|$.
In this situation the Fermi level only crosses
$\varepsilon_{K^{'}\downarrow}$, so the current is entirely
carried by the spin down electrons near the $K^{'}$ point.

\begin{figure}
\onefigure[width=7.0cm,height=5.0cm,angle=0]{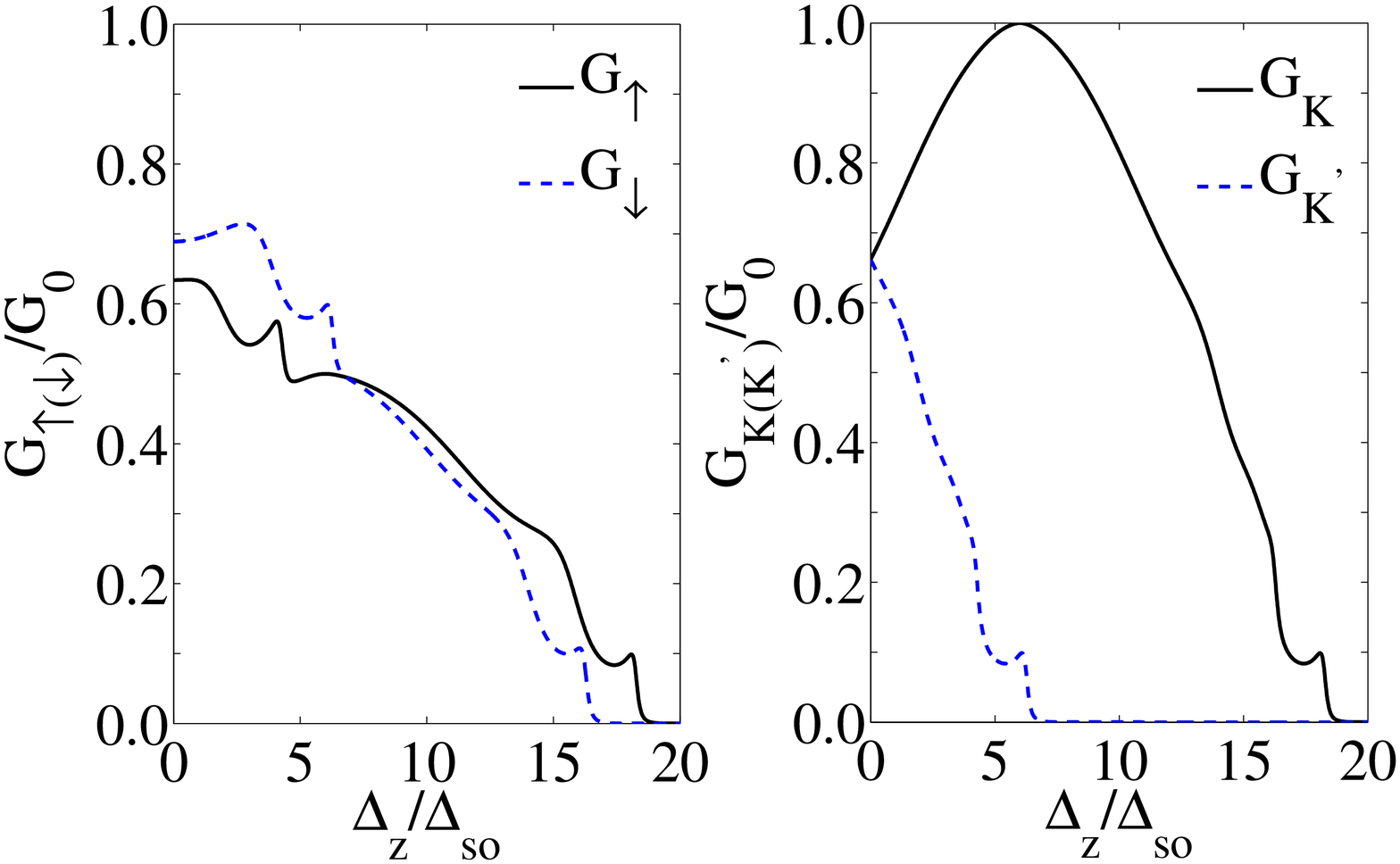}
\caption{(a) Spin-resolved conductance and (b) Valley-resolved
conductance as a function of $\Delta_{z}/\Delta_{so}$. The other
parameters are $\varepsilon_{F}=12\Delta_{so}$,
$\Delta_{\Omega}=6\Delta_{so}$ and $L=100~nm$.} \label{Fig06}
\end{figure}

Similarly, the dependence of $G_{K}$ and $G_{K^{'}}$ on
$\Delta_{\Omega}/\Delta_{so}$ (in the right panel of fig.
\ref{Fig05}) can be explained. If $\Delta_{\Omega}$ becomes larger
than $8\Delta_{so}$, happened when the Fermi level locates inside
the gap of both $\varepsilon_{K^{'}\uparrow}$ and
$\varepsilon_{K^{'}\downarrow}$ energy bands, the current is
entirely carried by the electrons near $\mathbf{K}$ point. So, the
current becomes fully valley polarized. Fully spin and valley
polarization can \textit{also} be obtained \textit{by changing the
perpendicular electric field }when $\Delta_{\Omega}\neq0$ as shown
in fig. \ref{Fig06}, in which we have plotted spin-resolved (left
panel) and valley-resolved (right panel) conductances as functions
of $\Delta_{z}/\Delta_{so}$.

%

\begin{figure}
\onefigure[width=7.0cm,height=5.0cm,angle=0]{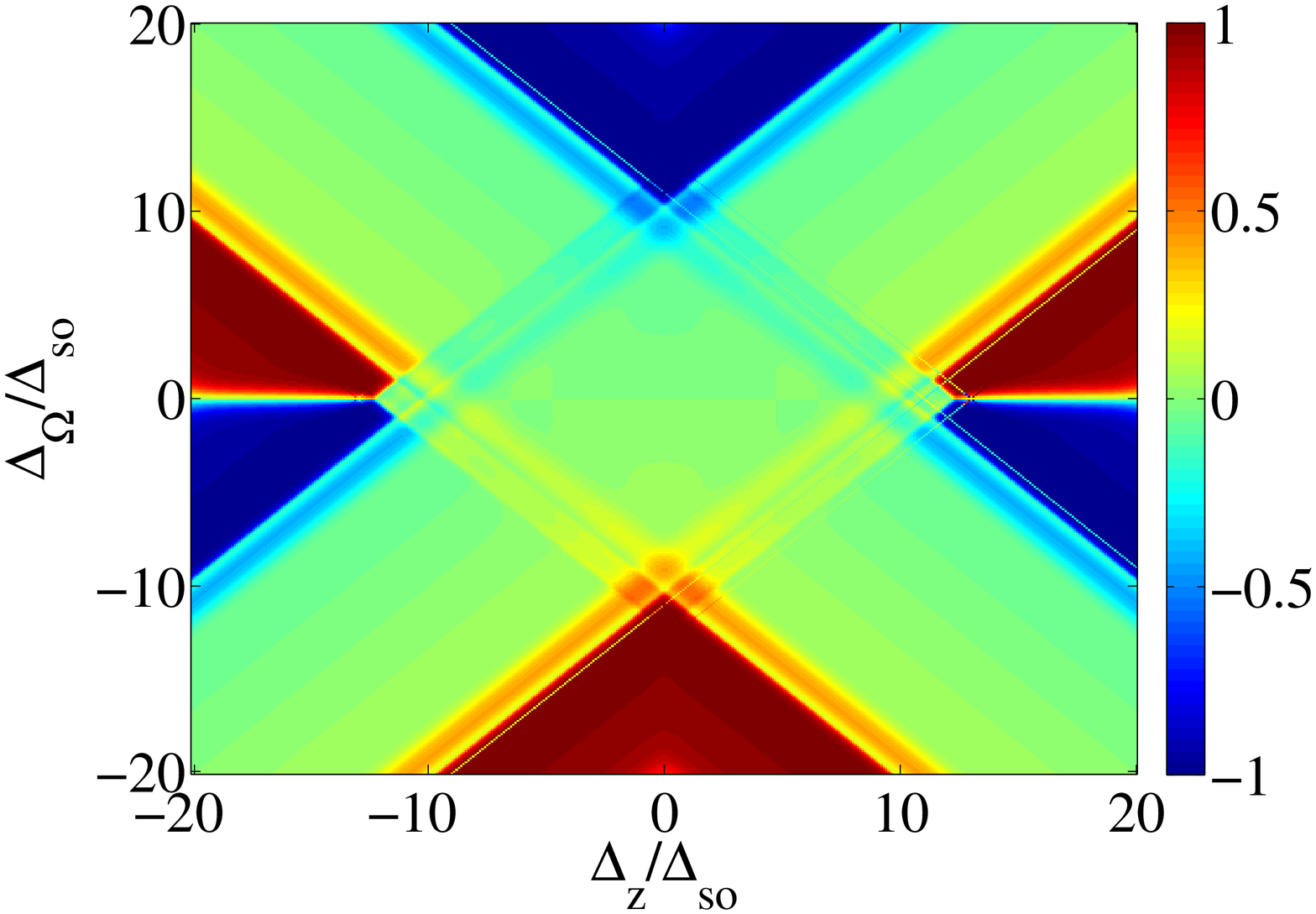}
\caption{Counter plot of spin polarization, $P_{s}$, as functions
of $\Delta_{\Omega}/\Delta_{so}$ and $\Delta_{z}/\Delta_{so}$. The
other parameters are $\varepsilon_{F}=12\Delta_{so}$ and
$L=100~nm$.} \label{Fig07}
\end{figure}

\begin{figure}
\onefigure[width=7.0cm,height=5.0cm,angle=0]{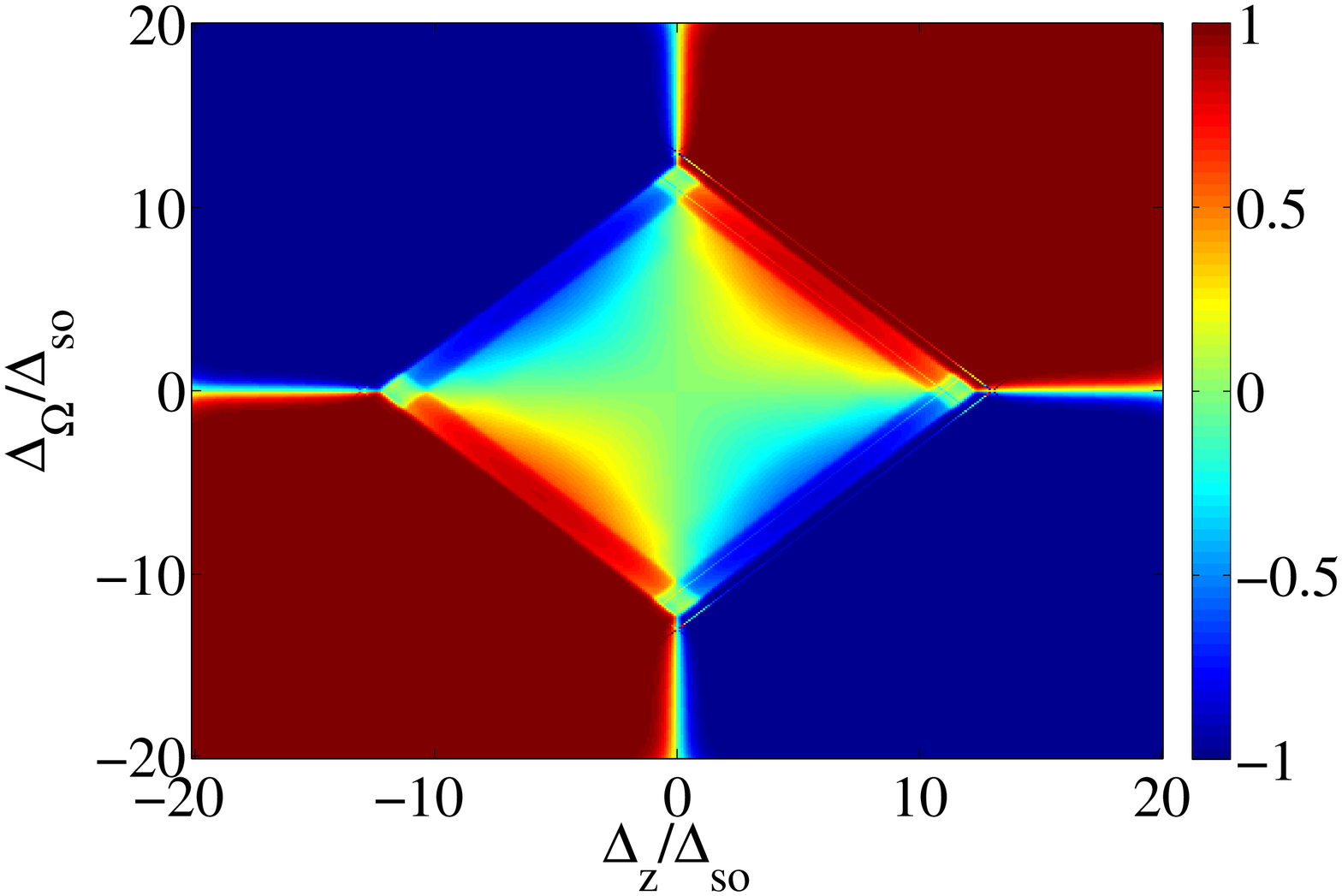}
\caption{Counter plot of valley polarization, $P_{v}$, as
functions of $\Delta_{\Omega}/\Delta_{so}$ and
$\Delta_{z}/\Delta_{so}$. The other parameters are
$\varepsilon_{F}=12\Delta_{so}$ and $L=100~nm$.} \label{Fig08}
\end{figure}

In the presence of the perpendicular electric field, the spin and
the valley polarization can be inverted by changing the direction
of the applied electric field (determined by the sign of
$\Delta_{z}$) and also by changing the polarization of the light
(determined by the sign of $\Delta_{\Omega}$). This is clear from
figs. \ref{Fig07} and \ref{Fig08}, in which we have shown the
counter plot of $P_{s}$ and $P_{v}$ as functions of
$\Delta_{\Omega}/\Delta_{so}$ and $\Delta_{z}/\Delta_{so}$.
$P_{s}$ is even in $\Delta_{z}$ but odd in $\Delta_{\Omega}$, so
we can invert $P_{s}$ only by reversing the circular polarization
of the light. While $P_{v}$ is odd in $\Delta_{z}$ and
$\Delta_{\Omega}$ except when $\Delta_{z}=0$ or
$\Delta_{\Omega}=0$. Hence, $P_{v}$ can be inverted either by
reversing the perpendicular electric field or by reversing the
circular polarization of the light. Figures \ref{Fig07} and
\ref{Fig08} also show the conditions necessary to realize a fully
spin or valley polarization. Form these figures, also, it is
evident that a fully valley polarized current can be attained in
wide a region of $\Delta_{\Omega}$ and $\Delta_{z}$. While for the
fully spin polarized current it isn't so. Further, these figures
show that the fully spin and valley polarizations can be obtained
together or separately.


\section{Summary and conclusions}

In summary, we studied ballistic transport of Dirac electrons
through a strip in silicene, when the strip is exposed to
off-resonant circularly polarized light and an electric field
applied perpendicular to the silicene plane. We showed that the
spin- and valley-conductance through the strip is polarized. This
can be understood by the spin-valley coupling in silicene via a
strong spin-orbit interaction, and modification of its band
structure by photon dressing process and the perpendicular
electric field. The spin (valley) polarization of the current can
be enhanced by tuning the light intensity and the value of the
perpendicular electric field, leading to perfect spin (valley)
filtering for certain of their values. Further, the spin (valley)
polarization can be inverted by reversing the perpendicular
electric field (by reversing the perpendicular electric field or
reversing the circular polarization of the light irradiation).

\end{document}